\documentclass[%
preprint,
prf,
 amsmath,amssymb,
 aps,
 prf,
]{revtex4-2}
\usepackage{graphicx} 
\usepackage{tikz}
\usetikzlibrary{positioning, arrows.meta}
\usepackage{subcaption}
\usepackage{soul}
\usepackage{listings}
\usepackage{xcolor}
\usepackage{booktabs}
\usepackage{multirow}
\usepackage{array}
\usepackage{geometry} 
\usepackage{xcolor}
\usepackage{float}    
\usepackage{makecell} 
\usepackage{footnote}
\begin{document}

\title{OpenFOAMGPT: a RAG-Augmented LLM Agent for OpenFOAM-Based Computational Fluid Dynamics}
\author{Sandeep Pandey}
\thanks{These authors contributed equally to this work.}
\affiliation{
Institute of Thermodynamics and Fluid Mechanics, Technische Universität Ilmenau, Ilmenau D-98684, Germany
}%

\author{Ran Xu}
\thanks{These authors contributed equally to this work.}
\affiliation{
Cluster of Excellence SimTech, University of Stuttgart, Stuttgart, Germany
}%

\author{Wenkang Wang}
\affiliation{
International Research Institute for Multidisciplinary Science, Beihang University, 100191 Beijing, China
}%

\author{Xu Chu}
\email{x.chu@exeter.ac.uk}
\affiliation{
Faculty of Environment, Science and Economy, University of Exeter, Exeter EX4 4QF, United Kingdom
}%
\affiliation{
Cluster of Excellence SimTech, University of Stuttgart, Stuttgart, Germany
}%

\begin{abstract}

This work presents a large language model (LLM)-based agent \textbf{OpenFOAMGPT} tailored for OpenFOAM-centric computational fluid dynamics (CFD) simulations, leveraging two foundation models from OpenAI: the GPT-4o  and a chain-of-thought (CoT)–enabled o1 preview model. Both agents demonstrate success across multiple tasks. While the price of token with o1 model is six times as that of GPT-4o, it consistently exhibits superior performance in handling complex tasks, from zero-shot case setup to boundary condition modifications, turbulence model adjustments, and code translation. Through an iterative correction loop, the agent efficiently addressed single- and multi-phase flow, heat transfer, RANS, LES, and other engineering scenarios, often converging in a limited number of iterations at low token costs. To embed domain-specific knowledge, we employed a retrieval-augmented generation (RAG) pipeline, demonstrating how preexisting simulation setups can further specialize the agent for sub-domains such as energy and aerospace. Despite the great performance of the agent, human oversight remains crucial for ensuring accuracy and adapting to shifting contexts. Fluctuations in model performance over time suggest the need for monitoring in mission-critical applications. Although our demonstrations focus on OpenFOAM, the adaptable nature of this framework opens the door to developing LLM-driven agents into a wide range of solvers and codes. By streamlining CFD simulations, this approach has the potential to accelerate both fundamental research and industrial engineering advancements.

\end{abstract}

\maketitle

\newpage
\section{Introduction}

Recently, the field of fluid mechanics has increasingly adopted data-driven methodologies, propelled by abundant high-fidelity simulation data and rapid advancements in machine learning \citep{Duraisamy.2019,yang2024data,pandey2020perspective,wang2024optimized,cremades2025additive,vinuesa2022enhancing}. These approaches have been developed to model turbulence both with and without governing equations \citep{wu2018physics,ling2016reynolds,Beck.2019,chu2024non,yang2019predictive,beck2023toward}, as well as to address complex heat transfer problems \citep{Chang.2018,Chu.2018b}. In addition, machine learning techniques have supported experimental measurements \citep{vinuesa2023transformative} and facilitated scientific discovery, for instance by enabling causal inference in fluid flows \citep{Wang.2021,wang2022spatial,liu2023interfacial}.

More recently, Large Language Models (LLMs) \citep{achiam2023gpt} are rapidly emerging as powerful tools in scientific and engineering domains, offering unprecedented capabilities in natural language understanding \citep{min2023recent}, automated reasoning \citep{ma2024llm}, and decision-making. For research and engineering, these models hold the potential to transform conventional workflows by enabling more efficient problem-solving \citep{song2023pre,wang2023prompt}, advanced optimization \citep{huang2024crispr,ma2024llm}, and accelerated scientific discovery \citep{chibwe2024evaluating,ramos2024review}. As recent studies demonstrate, LLMs can serve as intelligent assistants that both enhance traditional methodologies and pave the way for novel approaches to analysis, design, and simulation. \citet{buehler2024mechgpt,ni2024mechagents} presented MechGPT, a fine-tuned large language model designed to unify disparate knowledge domains—such as mechanics and biology—within the context of multiscale materials failure. By leveraging Ontological Knowledge Graphs for interpretability, retrieval-augmented generation, and expanded context lengths, MechGPT supports interdisciplinary insight, hypothesis generation, and the integration of new data sources.
\citet{ghafarollahi2024atomagents} introduced a physics-aware generative AI platform, AtomAgents, which autonomously tackles multi-objective materials design by combining LLM with multiple AI agents operating in a dynamic environment.
Moreover, \citet{ghafarollahi2024rapid} proposed a multi-agent AI model for automated metallic alloy discovery, integrating LLM-driven reasoning and planning, specialized AI agents for domain expertise, and a graph neural network for rapid retrieval of key physical properties, thereby accelerating complex materials design tasks.
In addition, \citet{buehler2024cephalo} introduced Cephalo, a family of multimodal vision LLM for materials science, combining visual and linguistic data to interpret complex scenes, generate precise textual descriptions, and enable advanced design workflows.

A prominent application of LLMs in fluid mechanics lies in their ability to facilitate equation discovery. For instance, \citet{du2024large} have demonstrated how LLMs can autonomously extract governing equations from data, capturing complex nonlinear relationships without extensive human intervention. 
Beyond equation discovery, LLMs have also been employed to streamline shape optimization problems. \citet{zhang2024usinglargelanguagemodels} for example, introduced a framework that leverages these models for optimizing geometric profiles, such as airfoils or axisymmetric configurations, to achieve reduced drag. 
Emerging research extends these advancements into areas such as microfluidics \cite{xu2024trainingmicrorobotsswimlarge}, where LLMs assist in decision-making for robotic motion planning under fluid flow constraints or guide microscale swimmers through complex fluidic environments. 
\citet{zhu2024fluid} introduces a framework that combines LLM with spatiotemporal-aware encoders to predict unsteady fluid dynamics, significantly improving accuracy and efficiency compared to traditional CFD methods. By leveraging pre-trained LLMs and graph neural networks, the model effectively integrates spatial and temporal information, achieving superior performance in long-horizon predictions for fluid datasets like airfoil and cylinder flows.
\citet{kim2024chatgpt} evaluated the potential of ChatGPT-generated MATLAB code guided by narrative-based prompts for geotechnical engineering applications, including seepage flow analysis, slope stability assessment, and X-ray tomography image processing. Although ChatGPT cannot fully replace conventional programming, it effectively refines code, minimizes syntax errors, and provides a logical starting framework when carefully directed by domain-specific guidance.
Equally significant are the applications of LLMs in automating and orchestrating CFD simulations. \citet{chen2024metaopenfoam} proposed an LLM-based multi-agent system capable of automating CFD workflows through natural language interactions. By employing retrieval-augmented generation (RAG), this platform can identify and correct potential errors, substantially lowering the technical barriers typically associated with CFD simulations and making these tools more accessible to non-specialists.

These advances illustrate how LLMs are reshaping the landscape of scientific research including fluid mechanics, from equation discovery and shape optimization to the automation of complex CFD workflows. In this work, we introduce an LLM-based, RAG-augmented agent for OpenFOAM, aiming to bridge the gap by providing a robust, conversational interface that can both retrieve domain-specific information and generate contextually informed instructions. With this LLM-based agent for CFD, our goal is to automate the workflow of performing simulations, significantly lower the expertise threshold, and thus boost overall productivity. Beyond OpenFOAM, we seek to illustrate that this flexible framework can be adapted to a wide range of other solvers and codes.

\section{Methodology}

OpenFOAM (Open Source Field Operation and Manipulation) is a widely utilized, open-source CFD solver package \citep{weller1998tensorial}. It offers a diverse collection of solvers for a broad range of applications, including incompressible \citep{Pandey.2017,Pandey.2018,Chu.2018a} and compressible flows \citep{liu2024simulation}, heat transfer \citep{evrim2021flow,Evrim.2020,Chu.2016c,Chu.2019,li2024microscale}, and multiphase systems \citep{liu2021simulation,liu2023large,yang2020droplet}. Unlike commercial CFD packages, OpenFOAM’s fully open-source architecture grants users the flexibility to modify existing solvers, develop new physical models. By relying on a robust C++ object-oriented design, it ensures maintainability, extensibility, and high parallel efficiency, making it a standard platform for both academic research and industrial engineering workflows. Its proven accuracy, adaptability, and scalability have made OpenFOAM an ideal environment for coupling with LLM-based agents that can interactively guide users through simulation setup, execution, and analysis. The version we use here is OpenFOAM-v2406 release.

GPT (Generative Pre-trained Transformer) is a family of AI foundamtion model developed by OpenAI that generates human-like text by predicting the next word based on the context of previous words. It is already widely used for tasks like chatbots, content creation, and code generation. We develop the OpenFOAM specific agent based on two foundation models of OpenAI: ChatGPT-4o (4o) \citep{achiam2023gpt} and ChatGPT-o1 preview (o1), each accessed through the OpenAI API. While GPT-4 is widely recognized for its robust language understanding, o1 \citep{wei2022chain,zhong2024evaluation} distinguishes itself through the chain-of-thought (CoT) mechanism, which enables better reasoning and more detailed problem-solving steps.
The API token pricing highlights a cost difference: the 4o model is priced at \$2.50 per million input tokens and \$10 per million output tokens, while the o1 model costs \$15.00 per million input tokens and \$60 per million output tokens.


\begin{figure}
    \centering
    \includegraphics[width=0.7\linewidth]{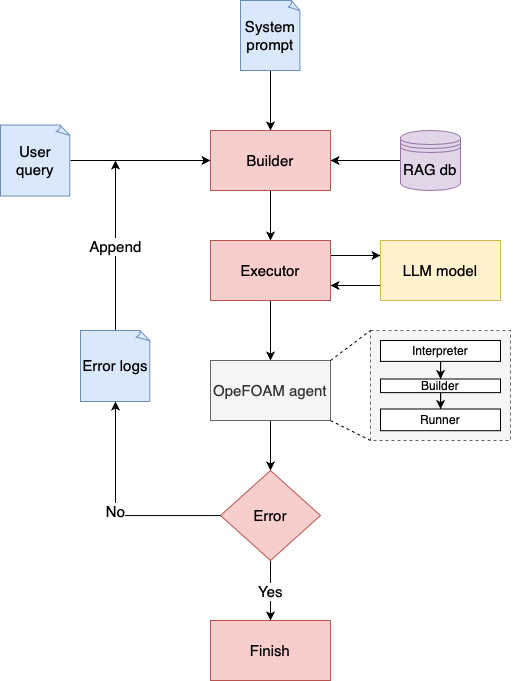}
    \caption{The design of the agent structure}
    \label{fig:agent}
\end{figure}


Figure~\ref{fig:agent} depicts the current, multi-layered structure of our agent \textbf{OpenFOAMGPT}. At the top, a system prompt is combined with a user query. The Builder module then interprets these instructions—consulting the RAG database as needed for domain-specific information—and transforms them into a structured plan. Next, the Executor orchestrates the workflow: it can direct the query to the LLM model for further reasoning or hand off tasks to the OpenFOAM agent for simulation-specific actions. Inside the OpenFOAM agent, an Interpreter, Builder, and Runner collaborate to set up and execute the necessary OpenFOAM operations. Throughout the simulation, error logs are monitored; if a failure is detected, the error data is appended to the user query and the process iterates. Otherwise, the workflow concludes successfully.
To ensure domain-specific completeness, the agent uses retrieval-augmented generation (RAG) when constructing or refining its plans. A dedicated RAG database—derived from OpenFOAM tutorial case descriptions—provides relevant details such as solver names, case names, and flow types. These descriptions are embedded into 1536-dimensional vectors via OpenAI's text-embedding-3-small model, allowing the agent to look up or incorporate analogous cases from public or private repositories seamlessly.

The addition of domain knowledge through RAG (Figure \ref{fig:RAG}) leads to a better prompting and accurate results \citep{lewis2020retrieval}. It is implemented through the LangChain framework \citep{langchain}. Although a LLM can reason about general knowledge, its performance in specialized tasks improves significantly when supplemented by domain-specific resources. By querying an indexed library of OpenFOAM tutorials, the RAG layer provides the LLM with validated best practices and updated methodologies. This additional context helps bridge typical knowledge gaps inherent to pre-trained models, ensuring that generated responses remain current with community standards in the field of CFD. Moreover, by further extending the knowledge base to include specialized topics—such as aerodynamics, process engineering, and heat transfer design, our general-purpose CFD agent can be seamlessly adapted to serve specific sub-disciplines. This modular structure not only maintains flexibility but also enables deep, domain-focused expertise, leading to more accurate and comprehensive solutions across a variety of fluid flow applications.

\usetikzlibrary{shapes,shapes.misc,shapes.geometric}

\begin{figure}[H]
    \centering
    \scalebox{0.7}{%
    \begin{tikzpicture}[font=\sffamily, 
                        line width=1pt, 
                        node distance=2.5cm,
                        >={Latex}]
    \tikzset{
        arrowstyle/.style={->, thick, draw=black},
        labelstyle/.style={font=\footnotesize},
    }
    \newcommand{\userfigure}{%
        \begin{tikzpicture}[scale=1]
            \draw[thick] (0,0) circle (0.3);
            \draw[thick] (-0.5,-0.8) arc[start angle=180, end angle=0, radius=0.5];
        \end{tikzpicture}
    }
    
    \node (userL) {\userfigure};
    \node[below=0.1cm of userL, font=\footnotesize, align=center]{User\\(Asks Question)};
    
    \node[ellipse, draw, thick, fill=green!10, 
          minimum width=2.3cm, minimum height=1.2cm, 
          align=center, right=3cm of userL](engine){Engine\\(Builds Prompt)};
    
    \node[cloud, cloud puffs=12, draw, thick, fill=red!10, 
          minimum width=2.5cm, minimum height=1.5cm, 
          align=center, right=3cm of engine](llm){LLM\\(Generates Response)};
    
    \node[right=3cm of llm](userR){\userfigure};
    \node[below=0.1cm of userR, font=\footnotesize, align=center]{User\\(Receives Response)};
    
    \node[draw, thick, fill=gray!10, 
          minimum width=3cm, minimum height=1.5cm,
          below=2.2cm of engine](store){};
    
    \begin{scope}[shift={(store.south west)}, x=1cm, y=1cm]
        \draw[thick] (0.2,0.3) -- (2.8,0.3);
    
        \draw[fill=black!20] (0.4,0.3) rectangle (0.6,0.8);
        \draw[fill=black!20] (0.8,0.3) rectangle (1.0,0.6);
        \draw[fill=black!20] (1.2,0.3) rectangle (1.4,0.7);
        \draw[fill=black!20] (1.6,0.3) rectangle (1.8,0.5);
        \draw[fill=black!20] (2.0,0.3) rectangle (2.2,0.9);
        \draw[fill=black!20] (2.4,0.3) rectangle (2.6,0.65);
    \end{scope}
    
    \node[font=\footnotesize, align=center] at ([yshift=-0.4cm]store.south) {Retrieval Library};
    \draw[arrowstyle] (userL) -- node[labelstyle,above]{Question} (engine);
    \draw[arrowstyle] (engine) --node[labelstyle,above]{Full Prompt}  (llm);
    \draw[arrowstyle] (llm) -- node[labelstyle,above]{Response} (userR);
    
    \draw[arrowstyle] ([xshift=-0.4cm]engine.south) 
      to[out=270,in=90] node[labelstyle,left]{Retrieval Query} ([xshift=-0.4cm]store.north);
    
    \draw[arrowstyle] ([xshift=0.4cm]store.north) 
      to[out=90,in=270] node[labelstyle,right]{Retrieved Texts} ([xshift=0.4cm]engine.south);
    \end{tikzpicture}
    }
    \caption{The structure of RAG}
    \label{fig:RAG}
\end{figure}
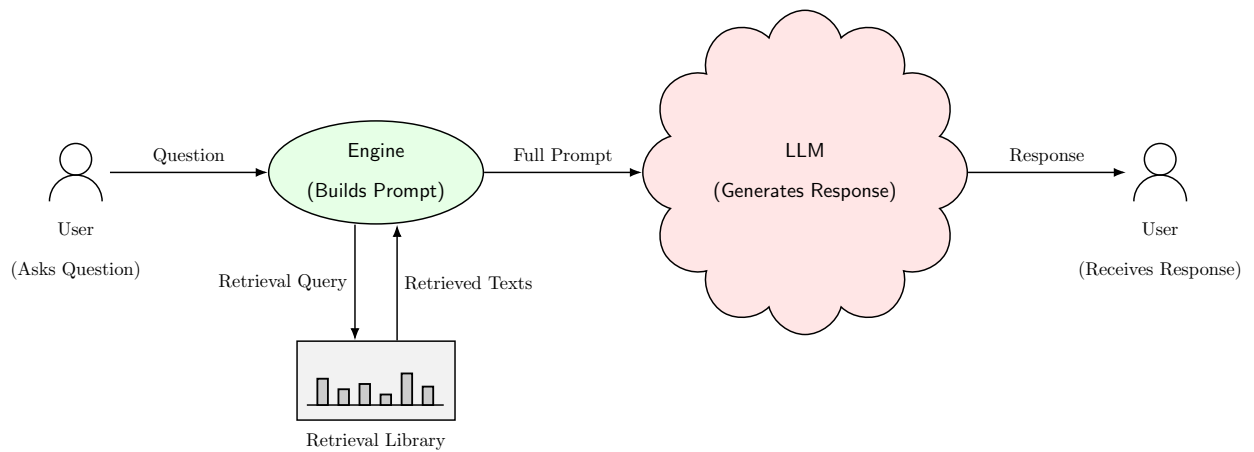

\section{Evaluations}

\subsection{Performance of zero-shot prompting}

It remains uncertain whether the developed agent \textbf{OpenFOAMGPT} can design and run simulations without further training. To evaluate the its capabilities, we selected six cases from the OpenFOAM tutorials encompassing single- and multiphase flow, laminar and turbulent regimes, and ranging from straightforward to application-oriented examples.

\begin{itemize}
    \item Cavity flow: This case simulates a laminar, isothermal, incompressible flow in a two-dimensional square domain using the icoFoam solver. The geometry consists of a square with all boundaries as walls. The top wall moves in the $x$-direction at a speed of 1 m/s, while the other three walls are stationary.  
    
    \item PitzDaily: It simulates incompressible, turbulent flow through the PitzDaily geometry using the k-$\epsilon$ model and the simpleFoam solver. The geometry is a sudden expansion channel in a two-dimensional plane, with a narrow entrance and a wider section after the sudden expansion. The entire channel forms a two-dimensional rectangular structure. 
    
    \item Hotroom: It is a turbulent natural convection inside a tall cavity which is simulated using the k-$\epsilon$ model and the buoyantBoussinesqSimpleFoam solver. The cavity has a two-dimensional rectangular shape with walls as boundaries. The left and right walls are adiabatic, the top wall is cold, and the bottom wall is hot. 
    
    \item Dambreak: It represents a simplified dam break with laminar flow, simulated using the VOF-based multi-phase solver (interFoam). The tank is a two-dimensional square, with a rectangular obstacle located at the center of the lower boundary. The top boundary is open, while the other three boundaries are stationary walls. The setup consists of a column of water at rest behind a membrane on the left side of the tank. At time $t$=0 s, the membrane is removed, causing the water to collapse. As the water impacts the obstacle, a complex flow structure is generated, including several pockets of trapped air. 
    
    \item Particle column: This case simulates the distribution and dynamic evolution of particle flow in a two-dimensional column using the MPPICFoam solver. The column is a rectangular domain, with an inlet at the top and an outlet at the bottom. The fluid is modeled as an incompressible, single-phase flow governed by the Navier-Stokes equations, while the particle dynamics are modeled using the Lagrangian tracking method. The multiphase particle dynamics model (MPPIC) is employed to account for particle collisions and contact stress. Forces acting on the particles include fluid drag, collision forces, and gravity. 
    
    \item Mixed vessel: This case simulates fluid flow in a rotary mixer or agitator using the pimpleFoam solver. The geometry is a two-dimensional cylinder with four rectangular barriers evenly distributed on both the outer and inner walls. The outer wall is stationary, while the inner wall rotates, with its motion simulated using a moving grid. 
    
\end{itemize}

Zero-shot prompting attempts to generate desired output based upto limited instructions without any example. As LLMs are trained on vast datasets, therefore, it often results in desired outputs. However, if underlying problem is difficult then model might not provide desired output. In this subsection, we analyze the performance of LLM models with zero-shot prompting.

\begin{table}[H]
    \centering
    \scriptsize
    \renewcommand{\arraystretch}{1.2} 
    \setlength{\tabcolsep}{5pt}       
    \begin{tabular}{|c|p{3.9cm}|p{4.7cm}|p{5.5cm}|} 
     \hline
     Case & gpt-4o & o1-mini & o1-preview \\ 
     \hline
     Cavity flow &  $\checkmark$ & $\checkmark$ & $\checkmark$ \\ 
     \hline 
     PitzDaily   &  $\times$ (\texttt{blockMeshDict} incorrect) & $\times$ (\texttt{blockMeshDict} incorrect) & $\times$ (\texttt{blockMeshDict} incorrect) \\ 
     \hline 
     Hotroom    & $\times$ (\texttt{blockMeshDict} incorrect) & $\times$ (\texttt{transportProperties} incorrect) &  $\checkmark$ \\ 
     \hline 
     Dambreak   & $\times$ (\texttt{blockMeshDict}incorrect) & $\times$ (\texttt{blockMeshDict} incorrect) &   $\times$ (\texttt{blockMeshDict} incorrect) \\ 
     \hline 
     Particle column &  $\times$ (\texttt{0} incorrect) & $\times$ (\texttt{0} incorrect)  & $\times$ (\texttt{kinematicCloudProperties} incorrect) \\ 
     \hline 
     Mixed vessel  &  $\times$ ( \texttt{blockMeshDict} incorrect) & $\times$ (\texttt{blockMeshDict} incorrect)  &  $\times$ (\texttt{blockMeshDict} incorrect) \\ 
     \hline
    \end{tabular}
    \caption{Evaluation zero-shot prompting of the agent with OpenAI's gpt-4o, o1-mini, and
    o1-preview.}
    \label{4o}
\end{table}

Table~\ref{4o} compares the performance of \textbf{OpenFOAMGPT} using foundation models 4o and o1, without RAG. When prompted solely with natural language, the agent employing model 4o can only produce simple two-dimensional square geometries (e.g., Cavity flow). Changing the model to o1-mini shows no improvement. In contrast, the agent using model o1-preview generates more complex geometries (e.g., Hotroom), although relying entirely on natural language is not ideal for specifying intricate shapes. Providing a \texttt{blockMeshDict} file as part of the input is therefore more effective.
After supplying a detailed \texttt{blockMeshDict} file, the agent with o1 demonstrates improved performance: even without RAG, it successfully handles the Cavity flow, PitzDaily, Hotroom, Dambreak, and Mixed vessel cases. However, for more complex problems such as Particle column—requiring unidirectional flow combined with particle flow—the agent encounters an unsolvable error when relying on the LLM alone.

\subsection{Few-Shot Prompting with RAG}

\begin{table}
    \centering
    \footnotesize
\begin{tabular}{|c|c|c|c|c|c|c|} 
 \hline
 \multirow{2}[0]{*}{case} &  \multicolumn{3}{c|}{4o} & \multicolumn{3}{c|}{o1-preview} \\
 \cline{2-7}
  &  Iterations &Token count & Cost (\$) & Iterations &Token count & Cost (\$)\\
\hline
    Cavity flow &2 & 6632 & 0.0327 & 1 & 9236 & 0.3956\\
\hline
    PitzDaily &4 & 19941 & 0.0808 & 7 & 40214 & 0.9069\\
\hline
    Hotroom & 9 & 37268 & 0.1269 & 5 & 36754 & 0.93795\\
\hline
    Dambreak & 5 & 22572 & 0.0853 & 8 & 20755 & 0.6294\\
\hline
    Particle column & 10 & 96666 & 0.3645 & 5 & 77157 & 1.6475\\
\hline
    Mixed vessel & 6 & 72769 & 0.2744 & 9 & 42750 & 1.0148\\
\hline
\end{tabular}
\caption{Comparison of the agent (based on 4o and o1-preview) with RAG and without RAG.}
\label{RAG}
\end{table}

\begin{figure}
    \centering
    \includegraphics[width=0.9\linewidth]{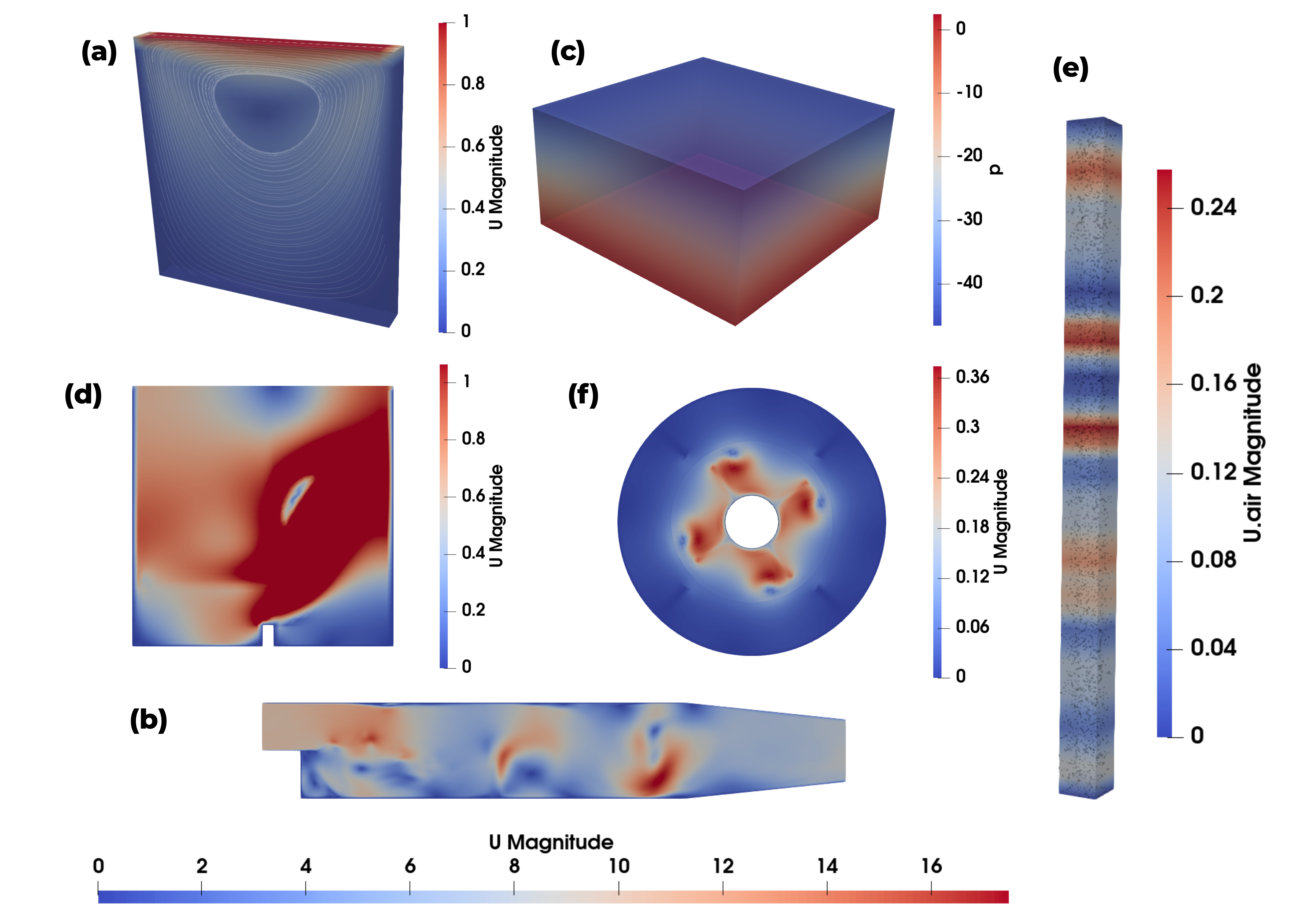}
    \caption{Case simulation results. (a) Cavity flow (b) PitzDaily (c) Hotroom (d) Dambreak (e) Particle column (f) Mixed vessel.}
    \label{result}
\end{figure}

In the previous subsection, \textbf{OpenFOAMGPT} based on GPT-4 (4o) without RAG capabilities succeeded only in the simplest cavity flow case, failing to set up more complex simulations. Here, we evaluate the improvement gained by enabling retrieval-augmented generation (RAG), with a summary of results presented in Table~\ref{RAG}. Figure~\ref{result} illustrates the simulation outcomes at the last time step. But in the case of particle column, the output of \texttt{kinematicCloudProperties} is different from the file provided in RAG because the output file has a length limit. Once equipped with RAG, the agent successfully sets up all tested cases, underscoring the significance of retrieving relevant domain knowledge for complex CFD workflows.

For the simplest cavity flow, only two interaction rounds were required, consuming approximately 6.6 thousand tokens. More complicated tasks, such as the particle column case, demanded ten interaction rounds and used around 96 thousand tokens. Application-oriented setups generally generated higher token usage, particularly when multiple output files or specialized model inputs were involved. Despite these increases, the associated monetary cost with 4o model varied from \$0.03 to \$0.36 per test scenario, a number that remains vastly lower than typical labor expenses. Execution times were under ten minutes in each instance, largely dependent on the number of interaction rounds and the amount of reference data consulted, while creating input files and running the solver remain swift.
Switching from the 4o model to the o1 model does not yield the anticipated reduction in iterations or token usage. Because the o1 model has a higher token price, overall costs rise significantly—up to \$1.60 in the particle column case. Consequently, given the differences in token pricing, the 4o model appears to be a more cost-effective choice.



\subsection{Alternate simulation conditions: initial- and boundary conditions, mesh resolution, thermophysical properties}

\begin{table}[H]
    \centering
    \footnotesize
\begin{tabular}{|>{\centering\arraybackslash}m{0.17\linewidth}|
>{\centering\arraybackslash}m{0.35\linewidth}|
>{\centering\arraybackslash}m{0.3\linewidth}|
>{\centering\arraybackslash}m{0.085\linewidth}|>{\centering\arraybackslash}m{0.085\linewidth}|}
\hline
    Case  & \multicolumn{2}{c|}{Alternate condition} & gpt-4o & o1-preview \\
\hline
    \multirow{4}{*}{Cavity flow} & Velocity of top wall movement & 1m/s $\to$ 2m/s & $\checkmark$ &  \\
\cline{2-5}          & Velocity of top wall movement & 1m/s $\to$ $5\sin\left(2\pi \frac{t}{0.1}\right)$ & $\times$ & $\checkmark$ \\
\cline{2-5}          & Mesh resolution & 20$\times$20$\times$1 $\to$  $\times$15$\times$1 & $\checkmark$ &   \\
\cline{2-5}          & endTime  & 3 $\to$ 5 & $\checkmark$ &   \\
\hline
    PitzDaily & Speed of the inlet & 10m/s $\to$ 20m/s & $\checkmark$ &   \\
\hline
    Hotroom & Temperature of HOT\_WALL & 310K  $\to$ 320K & $\checkmark$ &   \\
\hline
    Dambreak & Liquid inside the membrane & water  $\to$ oil & $\checkmark$ &   \\
\hline
    \multirow{3}{*}{Particle column} & Velocity of the fluid and particles & 1m/s  $\to$ 2m/s & $\checkmark$ &   \\
    
\cline{2-5}          & Type of fluid & Air  $\to$ CO &  $\times$ & $\checkmark$ \\
\hline
    Mixed vessel & Angular speed of rotation & 20rad/s  $\to$ 15rad/s & $\checkmark$ &   \\
\hline
\end{tabular}%
\caption{Tasks of alternate initial- and boundary conditions.}
\label{BCIC}
\end{table}

In engineering research and design, it is often necessary to explore a broad parameter space by adjusting simulation conditions—such as initial and boundary conditions or thermophysical properties—ranging from simple, homogeneous, steady-state parameters to unsteady, heterogeneous scenarios. Table~\ref{BCIC} summarizes these tasks, showing that the agent with o1-preview model can successfully handle all changes. Even the agent using the 4o model can accomplish most of these objectives.

A noteworthy example involves assigning an unsteady boundary condition, \(5\sin\bigl(2\pi\, \frac{t}{0.1}\bigr)\), in the cavity flow setup. \ul{The prompt includes the requirement but no information of how it should be done.} 
In fact, \textbf{OpenFOAMGPT} chose the right boundary condition \texttt{codedFixedValue}, not others such as \texttt{fixedValue}, \texttt{oscillatingFixedValue}.
This operation is more complex than simply replacing a constant value, yet the OpenFOAMGPT with the foundation model o1 managed it by generating advanced OpenFOAM scripting (excerpted in Listing~\ref{dynamiccode}). This accomplishment illustrates o1’s capacity for high-level reasoning and planning, further emphasizing its potential for sophisticated CFD applications.

\begin{lstlisting}[caption={Generated code in the \texttt{U} file.}, label={dynamiccode}, basicstyle=\ttfamily\scriptsize, aboveskip=1pt, belowskip=1pt]
movingWall
{
    type            codedFixedValue;
    value           uniform (0 0 0);
    redirectType    sineVelocity;

    code
    #{
        scalar amplitude = 5.0; // Amplitude in m/s
        scalar period = 0.1;    // Period in seconds
        scalar omega = 2.0 * M_PI / period;
        vector velocity(amplitude * sin(omega * this->db().time().value()), 0, 0);
        operator==(velocity);
    #};
}

\end{lstlisting}

\subsection{Zero-Shot addition or replacement of turbulence models (RANS and LES)}


Many engineering simulations involve turbulent flows that require closure models for the Reynolds-averaged Navier–Stokes equations or subgrid-scale formulations in large eddy simulations. Numerous turbulence models are available—each with its own level of complexity, computational cost, and approach to flow anisotropy—making it routine for R\&D teams to test and compare different models. However, this process extends well beyond merely switching a model’s name, as it often necessitates adjusting/adding/deleting initial and boundary conditions for different transport equations, as well as selecting suitable numerical schemes.
In this subsection, we examine the zero-shot capability of \textbf{OpenFOAMGPT} in adding or swapping turbulence models across various cases. Our goal was to transition from the standard \(k\text{-}\epsilon\) model to other RANS models, including RNG \(k\text{-}\epsilon\), \(k\text{-}\omega\) SST, and the three-equation eddy-viscosity model $k-kl-\omega$ as well as the Launder–Reece–Rodi (LRR) Reynolds-stress turbulence model. It is worth noting that the $k-kl-\omega$ model appears less commonly used and is more complex. But the foundation model o1 performs effectively in this context (excerpted in Listing~\ref{turbulencecode}). For LES, we replaced the dynamicKEqn model with a standard Smagorinsky turbulence model. Table~\ref{model} summarizes these tasks and provides key observations. 
Although the agent successfully handled most scenarios, it encountered difficulties when applying a \(k\text{-}\epsilon\) model to the particle column case. The particle column scenario itself involves higher complexity, posing challenges. Nevertheless, our results underscore the agent’s potential to significantly streamline or even automate advanced CFD tasks with minimal human input.

\begin{lstlisting}[caption={Generated code in the \texttt{turbulenceProperties} file.}, label={turbulencecode}, basicstyle=\ttfamily\scriptsize, aboveskip=1pt, belowskip=1pt]
simulationType  RAS;
RAS
{
    RASModel        kkLOmega;
    turbulence      on;
    printCoeffs     on;
    kkLOmegaCoeffs
    {
        sigmaK       2.0;
        sigmaL       0.66666;
        alphaOmega   0.52;
        betaOmega    0.072;
    }
}

\end{lstlisting}

\begin{table}[H]
    \centering
    \footnotesize
\begin{tabular}{
    |>{\centering\arraybackslash}m{0.15\linewidth}|
    >{\centering\arraybackslash}m{0.5\linewidth}|
    >{\centering\arraybackslash}m{0.15\linewidth}|>{\centering\arraybackslash}m{0.15\linewidth}|
} 
\hline
Case  & Alternate turbulence models & gpt-4o & o1-preview \\
\hline
    \multirow{2}[0]{*}{Cavity flow} & kEpsilon$\to$ RNGkEpsilon/kOmegaSST/LRR & $\checkmark$ & \\
    \cline{2-4} 
     & kEpsilon $\to$ kkLOmega & $\times$ & $\checkmark$\\
\hline
    \multirow{2}[0]{*}{PitzDaily} & 
    kEpsilon$\to$ kOmegaSST & $\checkmark$ & \\ 
    \cline{2-4} 
 & dynamicKEqn (LES) $\to$ Smagorinsky (LES) & $\checkmark$ & \\

\hline
    Dambreak & Laminar flow $\to$ kEpsilon & $\checkmark$ &  \\
\hline
    Particle column & Laminar flow$\to$ kEpsilon & $\times$ & $\times$ (\texttt{kinematicCloudProperties} incorrect) \\
\hline
    Mixed vessel & Laminar flow $\to$ kEpsilon & $\times$ & $\checkmark$\\
\hline
\end{tabular}
\caption{Tasks of add/alternate turbulence models}
\label{model}
\end{table}

\subsection{Zero-Shot code translation with o1-preview}

Code translation is a frequent and essential operation in software development.  It is often necessary to migrate or port code between different versions or distributions  of the same framework. For instance, two distinct OpenFOAM distributions 
have been developed and maintained independently over the years:  \url{www.openfoam.com} and \url{www.openfoam.org}. 
These versions are generally incompatible with each other.  As a result, users and developers regularly face the need to import a case from one distribution to the other, which requires substantial modifications to source code and library configurations. In this subsection, we illustrate how \textbf{OpenFOAMGPT} with the o1 model can facilitate this cross-platform translation process.

\begin{table}[H]
    \centering
    \footnotesize
    \begin{tabular}{|c|c|c|}
        \hline
        Task & Web-based LLM (o1-preview)  & Agent with o1-preview \\
        \hline
        T-Junction (RANS )&  $\times$ functions for post-processing in \texttt{controlDict} & $\checkmark$ \\
        \hline
        Channel395  (LES)&  $\checkmark$ & $\checkmark$\\
        \hline
        PitzDaily (RANS)  &  $\times$ Incompatible dimensions for operation & $\checkmark$ \\
        \hline
        Motorbike (RANS)  &  $\times$  &  $\times$  \texttt{snappyHexMeshDict} incorrect\\
        \hline
    \end{tabular}
    \caption{Code translation from OpenFOAM 12 platform (\protect\url{www.openfoam.org}) to V2406 platform (\protect\url{www.openfoam.com}) }
    \label{translation}
\end{table}

In Table~\ref{translation}, we present the results of translating four tutorial cases  (T-Junction, Channel395, PitzDaily, and Motorbike) from the OpenFOAM 12 platform (\url{www.openfoam.org}) 
to the OpenFOAM V2406 platform (\url{www.openfoam.com}). \ul{During the translation process, we provided code only for the OpenFOAM 12 platform and did not include any information related to the OpenFOAM V2406 platform. When using the web-based LLM (o1-preview), Channel395 is the only case that translates successfully without inconsistencies.} The remaining three cases encountered various errors, ranging from inaccurate function entries in \texttt{controlDict} to dimension mismatches in the field operations. Such issues highlight the complexities of cross-platform code translation, emphasizing the need for meticulous validation and targeted debugging. However, when using the agent, only the Motorbike case experienced errors that prevented it from running, while the PitzDaily case failed to converge after a few steps. This observation underscores the effectiveness of using agent to resolve issues in similar situations, further emphasizing the necessity of its use.



\section{Conclusion, limitations and outlook}

We have presented a LLM-based agent \textbf{OpenFOAMGPT} for OpenFOAM-focused computational fluid dynamics, integrating two foundation models from OpenAI—namely, the widely used GPT-4 (4o) and a chain-of-thought (CoT)–enabled o1 model. Although o1 incurs about six times the token cost, it demonstrated a clear edge in handling complex tasks, underscoring its potential value in specialized CFD workflows. Through iterative correction loop, \textbf{OpenFOAMGPT} successfully tackled practical simulation tasks such as zero-shot case setup, modifications to initial and boundary conditions, zero-shot turbulence model alternation, zero-shot code translation, and more. Representative test scenarios spanned single- and multi-phase flow, heat transfer, RANS, LES, and various engineering cases, all of which were resolved within a limited number of iteration loops at low token expenses.

Despite the robust performance of the pre-trained LLMs—particularly the CoT-enabled o1 model—our findings highlight the importance of retrieval-augmented generation (RAG) for embedding domain-specific knowledge. By incorporating existing simulation setups into the RAG pipeline, the agent can be further specialized for sub-domains such as energy sector, aerospace, or any area requiring tailored CFD solutions. Nonetheless, a measure of human oversight remains critical to ensure correctness and adapt to evolving contexts. Furthermore, fluctuations in model performance over time serve as a reminder that users must carefully monitor this change (with no official reminder) updates for mission-critical or high-precision applications.

Looking ahead, continued enhancements in hardware and frequent release of new powerful models will likely lower token costs and further enhance reasoning capabilities of the foundation model. While our case study centered on OpenFOAM, the adaptability of this approach paves the way for integrating LLM-based agents into a broad spectrum of solvers and codes. By reducing barriers to high-fidelity simulations, our framework holds promise for accelerating innovation across both fundamental research and industrial engineering processes.

\bibliography{sn-bibliography}

\end{document}